\documentclass[proof]{pasj00}
\SetRunningHead{Enoki et al.}{Environments of Quasars}
\Received{2002/8/1}
\Accepted{2002/11/26}
\Published{}        

\begin{document}
\title{Relations Between Environments of Quasars and Galaxy Formation}
\author{Motohiro \textsc{Enoki} \altaffilmark{1,2}, 
Masahiro \textsc{Nagashima} \altaffilmark{2},
and Naoteru \textsc{Gouda} \altaffilmark{2}}

\altaffiltext{1}{Department of Earth and Space Science, Graduate School of Science, \\
Osaka University, Toyonaka, Osaka, 560-0043}

\altaffiltext{2}{National Astronomical Observatory, Osawa 2-21-1, Mitaka, Tokyo, 181-8588} 

\maketitle

\email{enoki@vega.ess.sci.osaka-u.ac.jp, masa@th.nao.ac.jp, naoteru.gouda@nao.ac.jp}

%%%%%%%%%%%%%%%%%%%%%%%%%%%%%%%%%%%%%%%%%%%%%%%%%%%%%%%%%%%%
\begin{abstract}
We investigate the environments of quasars such as number distribution of
 galaxies 
using  a semi-analytic
model which includes both galaxy and quasar formations based on the
hierarchical  clustering scenario. We assume that a supermassive
black hole is fueled by accretion of cold gas and that it is a source of
quasar activity during a  major merger of the quasar host galaxy with
another galaxy. This major merger causes spheroid formation of the
host galaxy. Our model can reproduce not only general form of the galaxy luminosity
 functions in the local Universe but also the observed relation between a supermassive black
hole mass and a spheroid  luminosity, the present black hole mass
 function and the quasar luminosity functions
at different  redshifts. Using this model, we predict the mean
 number of quasars per halo, bias parameter of quasars  and the probability
 distribution of the number of galaxies around quasars. In our model, 
 analysis of the mean number of quasars per  halo shows 
 that the spatial
distribution of galaxies is different from that of quasars. Furthermore,
we found from calculation of the probability distribution of galaxy
 numbers that at $0.2 \lesssim z \lesssim 0.5$, most quasars are likely
 to reside in galaxy groups. On the other hand, at $1 \lesssim z \lesssim 2$     
 most quasars seem to reside in more varied  environments than at a
 lower redshift; quasars reside in environments ranging from small groups of galaxies to
 clusters of galaxies. Comparing these predictions with
 observations in future will enable us to constrain our quasar formation model.  
\end{abstract}

\section{Introduction}\label{intro}
The environments of quasars provide important clues to the physical
processes of their formation and also yield important
information about the relations between the distribution of quasars and
the large-scale structure of the universe.
For more than three decades, we have known that quasars are
associated with enhancements in the spatial distributions of galaxies (\cite{BSG69}).
Studies of the environments of quasars in the nearby universe ($z
\lesssim 0.4$)
have shown that quasars reside in environments ranging from small to moderate groups of galaxies
rather than in rich clusters (e.g. \cite{BC91}; \cite{FBK96};
\cite{MD01}).
In order to interpret the observational results of the environments of
quasars at  low redshifts and predict the environments of
quasars at high redshifts, a physical model of quasar formation
based on cosmological context is required.      

 It has become widely accepted that quasars are fueled by accretion of
 gas onto  supermassive black holes (SMBHs)  in the nuclei of  host
 galaxies since \citet{Lyn} proposed this idea on
 quasars. Recent  observations of galactic centers suggest that a
 lot of nearby galaxies  have central black  holes and their estimated
 masses correlate with the  luminosities of
 spheroids\footnote{Throughout this paper, we refer to bulge or
 elliptical galaxy as {\it spheroid}.} of their
 host galaxies (e.g. \cite{KR95}; \cite{Mag98}; \cite{MF00}). 
The connection between SMBHs and their host spheroids
 suggests that the formation of SMBHs physically links the formation of
 the spheroids which harbor the SMBHs. Thus, this implies that the
 formation of quasars is 
 closely related to the formation of galaxies, especially of spheroids.   
 Therefore, in order to study the
 formation and evolution of quasars, it is necessary to construct a unified
 model which includes both galaxy formation and quasar formation.

Recently, some authors have tried to construct galaxy formation models
on the basis of the theory of hierarchical structure formation in  cold dark
matter (CDM) universe. These efforts are  referred to as  semi-analytic models (SAMs) of
galaxy formation.
In the CDM universe, dark matter halos cluster gravitationally
and merge together in a manner that depends on the adopted power spectrum of
initial density fluctuations.  In each of the merged dark halos,
radiative gas cooling, star formation, and  supernova feedback
occur.  The cooled dense gas and stars constitute {\it galaxies}.  These
galaxies sometimes merge together in a common dark halo and more massive
galaxies form. 
In SAMs, the merger trees of dark matter halos are constructed using a
Monte-Carlo algorithm and simple models are adopted to describe the above
gas processes. 
Stellar
population synthesis models are used to calculate the luminosities and
colors of model galaxies. It is therefore straightforward to
understand how galaxies form and evolve within the context of this
model.
 SAMs successfully have reproduced a variety of observed
features of local galaxies such as their luminosity functions, color
distribution, and so on (e.g. \cite{KWG93}; \cite{CL94}, \yearcite{CL00};
\cite{SP99}; \cite{NTGY01}, \yearcite{NYTG02}).

In these models, it is assumed that disk stars are  formed by cooling of gas in the halo.
If two galaxies of comparable mass merge, it is assumed that starbursts
 occur and  form the spheroidal component in the center of
the galaxy. $N$-body simulations have shown that a merger hypothesis for
the origin of spheroids can explain    
 their detailed internal structure (e.g.
\cite{Barn}; \cite{Her92}, \yearcite{Her93}; \cite{HHS94}). Kauffmann
and Charlot (\yearcite{KC98}) have demonstrated that the merger
scenario for the formation of elliptical galaxies is consistent with the color-magnitude
relation and its redshift evolution (see also \cite{NG01}).
On the other hand, hydrodynamical simulations have shown that a merger of
galaxies 
drives gas to fall rapidly to the center of a merged system and to fuel
nuclear starburst (\cite{NW83}; \cite{MH94}, \yearcite{MH96};
\cite{BH96}). Moreover, observed 
images of quasar hosts show that many quasars reside in interacting systems or elliptical galaxies (\cite{BKSS}). Therefore, it
has often been thought that the major merger of galaxies would be a
possible  mechanism for
quasar and spheroid formation.

So far, a lot of studies on quasar evolution based on the hierarchical
 clustering scenario have been carried out with the assumption
 that the formation of quasars is linked to the first collapse of dark
matter halos with galactic mass and that these models can explain the decline of quasar number
density at $z \gtrsim 3$ (e.g. \cite{ER88}; \cite{HR93}) and 
properties of luminosity functions of quasars (e.g. \cite{HL98}; \cite{HNR98}; \cite{HMKYU}).
 However, if quasars are directly linked to spheroids of host
 galaxies rather than to dark matter halos, the
 approximation of a one-to-one relation between quasar hosts and dark
 matter halos would be very crude, especially at low redshift. Therefore, it is
 necessary to construct a model related to spheroid formation and SMBH
 formation directly. Kauffmann and Haehnelt
 (\yearcite{KH00}) introduced a unified model
of the evolution of galaxies and quasars within the framework of SAM (see also
\cite{Cat01}). They assumed that SMBHs are formed and
fueled  during major galaxy mergers and  their model reproduces 
quantitatively the observed relation between spheroid luminosity and black hole
mass  in nearby galaxies, the strong  evolution of the quasar population with redshift,
and the relation between the luminosities of nearby quasars and those of
their host galaxies.  

In this paper, we investigate properties of quasar environments, using a
SAM
incorporated simple quasar evolution model. We assume that SMBHs are
formed and fueled  during major galaxy mergers  and the fueling
process leads quasar activity. While this assumption is similar to the model of  Kauffmann and Haehnelt
 (\yearcite{KH00}),
our galaxy formation model and the adopted model of fueling process are
different from their model.     
Here we focus on optical
properties of quasars and attempt to consider the number of quasars per
halo, effective bias parameter of quasars
 and the number
of galaxies around quasars as  characterizations of
environments of quasars, because a)  
these quantities provide a direct measure of bias in their
distribution with respect to galaxies and b) comparing results of the model
with observations will enable us to constrain our quasar formation model.

The paper is organized as follows: in \S \ref{model}  
we briefly review our SAM for galaxy formation; in \S \ref{qsomodel} we
introduce the quasar formation
model; in \S \ref{env} we calculate the galaxy number distribution function
around quasars; in \S \ref{disc} we provide a summary and discussion.

In this study, we use a low-density, spatially flat cold dark matter
($\Lambda$CDM) universe with
the present density parameter $\Omega_0=0.3$, the cosmological constant
$\lambda_0=0.7$, the Hubble constant in units of $100 {\rm km \ s^{-1}\ {Mpc^{-1}}}$ $h=0.7$
and the present rms density fluctuation in spheres of $8 h^{-1} {\rm
Mpc}$ radius $\sigma_8=1.0$.

\section{Model of Galaxy Formation}\label{model}
 In this section we briefly describe our SAM for the galaxy formation
 model, details of which are shown in  \citet{NTGY01}. Our
present SAM analysis obtains essentially the same results as the previous
SAM analyses, with minor differences in a number of details. 

First, we construct Monte Carlo realizations of merging histories of dark
 matter halos using the method of \citet{SK99}, which is 
 based on the extended 
Press-Schechter formalism (\cite{PS74}; \cite{Bow91};
\cite{BCEK}; \cite{LC93}).  We adopt the power spectrum for the
specific cosmological model from \citet{BBKS}. Halos with
circular velocity $V_{\rm circ}<$40km~s$^{-1}$ are treated as diffuse
 accretion matter.  The evolution of the baryonic component is followed
 until the output redshift coincides with the redshift interval of $\Delta z=0.06(1+z)$,
corresponding to the dynamical time scale of halos which collapse at
the redshift $z$. Note that \citet{SKSS} recently pointed out that a much
 shorter timestep is required to correctly reproduce the mass function
 given by the Press-Schechter formalism. However, a serious problem
 exists only at small mass scales ($\lesssim 10^{11} M_{\odot}$). Thus we
 use the above prescription of timestep.    

If a dark matter halo has no progenitor halos, the mass fraction of the
gas in the halo is given by $\Omega_{\rm b}/\Omega_{\rm 0}$ , where $\Omega_{\rm b}=0.015h^{-2}$ is the baryonic density
parameter constrained by primordial nucleosynthesis calculations
(e.g. \cite{SYB00}). Note that a recent measurement of the anisotropy of
the cosmic microwave background by the BOOMERANG project suggests a
slight higher value, $\Omega_{\rm b} \sim 0.02 h^{-2}$ (\cite{BOOM}).
\citet{CL00} have already investigated the effect of changing
$\Omega_{\rm b}$ and showed that this mainly affects the value of the
invisible stellar mass fraction such as brown dwarfs parameterized by
$\Upsilon$ (see below).   
When a dark matter halo collapses, the gas in the halo is shock-heated
to the virial temperature of the halo.  We refer to this heated gas as
the {\it hot gas}.  At the same time, the gas in dense regions of the
halo cools due to efficient radiative cooling.  We call this cooled
gas the {\it cold gas}.  Assuming a singular isothermal density distribution of
the hot gas and using the metallicity-dependent cooling function by
\citet{SD93}, we calculate the amount of cold gas which
eventually falls onto a central galaxy in the halo.  In order to avoid
the formation of unphysically large galaxies, the above cooling process
is applied only to halos with $V_{\rm circ}<$400 km~s$^{-1}$. This handling would be needed because the simple isothermal distribution
forms so-called ``monster galaxies'' due to the too efficient cooling at
the center of halos.  While this problem will probably solved by
adopting another isothermal distribution with central core (\cite{CL00}),
we take the above approach for simplicity.

Stars are formed from the cold gas at a rate of $\dot{M}_{*}={M_{\rm
cold}}/{\tau_{*}}$, where $M_{\rm cold}$ is the mass of cold gas and
$\tau_{*}$ is the time scale of star formation.  We assume that
$\tau_{*}$ is independent of $z$, but dependent on $V_{\rm circ}$ as
follows:

\begin{equation}
\tau_{*}=\tau_{*}^{0}\left(\frac{V_{\rm circ}}{300\mbox{km~s}^{-1}}\right)
^{\alpha_{*}}\label{eq:taustar}.
\end{equation}
The free parameters of $\tau_{*}^{0}$ and $\alpha_{*}$ 
are fixed by matching the observed mass fraction of cold gas in neutral
form in the disks of spiral galaxies. 
In our SAM, stars with masses larger than $10M_\odot$ explode as Type II
supernovae (SNe) and heat up the surrounding cold gas.  This SN feedback
reheats the cold gas to hot gas at a rate of $\dot{M}_{\rm
reheat}=\beta \dot{M}_{*}$, where $\beta$ is the efficiency of
reheating. We assume that $\beta$ depends on $V_{\rm circ}$ as
follows:
\begin{equation}
\beta=\left(\frac{V_{\rm circ}}{V_{\rm hot}}
\right)^{-\alpha_{\rm hot}}. \label{eq:feedback}
\end{equation}
The free parameters of $V_{\rm hot}$ and $\alpha_{\rm hot}$ 
are determined by matching the local luminosity function of galaxies.  
With these $\dot{M}_{*}$ and $\dot{M}_{\rm reheat}$ thus determined, we
obtain the masses of hot gas, cold gas, and disk stars as a function of
time during the evolution of galaxies. 

Given the star formation rate as a function of time, the absolute
luminosity and colors of individual galaxies are calculated using a
population synthesis code by \citet{KA97}. The
initial stellar mass function (IMF) that we adopt is the power-law IMF
of Salpeter form with lower and upper mass limits of $0.1$M$_{\odot}$
and $60$M$_{\odot}$, respectively.  Since our knowledge of the lower
mass limit is incomplete, there is the possibility that many brown
dwarf-like objects are formed.  Therefore, following \citet{CL94},
we introduce a parameter defined as $\Upsilon=(M_{\rm lum}+M_{\rm
BD})/M_{\rm lum}$, where $M_{\rm lum}$ is the total mass of luminous
stars with $m\geq 0.1M_\odot$ and $M_{\rm BD}$ is that of invisible
brown dwarfs. To account for extinction by internal dust we adopt a
simple model by \citet{WH96} in which the optical depth in $B$-band is related
to the luminosity as $\tau_{B}=0.8(L_{B}/1.3\times
10^{10}L_{\odot})^{0.5}$.  Optical depths in other bands are calculated
by using the Galactic extinction curve, and the dust distribution in
disks is assumed to be the slab model considered by \citet{SP99}. 

When several progenitor halos have merged, the newly formed larger
halo should contain at least two or more galaxies which had originally
resided in the individual progenitor halos.   We identify
the central galaxy in the new common halo with the central galaxy
contained in the most massive of the progenitor halos.  Other galaxies
are regarded as satellite galaxies.
These satellites merge by either dynamical friction or random collision.
The time scale of merging by dynamical friction is given by
\begin{equation}
\tau_{\rm fric}=\frac{260}{\ln\Lambda_{\rm c}}\left(\frac{R_{\rm H}}{\rm Mpc}\right)^{2}
\left(\frac{V_{\rm circ}}{10^{3}{\rm km~s}^{-1}}\right)
\left(\frac{M_{\rm sat}}{10^{12}M_{\odot}}\right)^{-1}{\rm Gyr},
\end{equation}
where $R_{\rm H}$ and $V_{\rm circ}$ are the radius and the circular
velocity of the new common halo, respectively, $\ln\Lambda_{\rm c}$ is
the Coulomb logarithm, and $M_{\rm sat}$ is the mass of the satellite galaxy
including its dark matter halo \citep{BT}.  When the
time passed after a galaxy becomes a satellite exceeds $\tau_{\rm
fric}$, a satellite galaxy infalls onto the central galaxy.  On the
other hand, the mean free time scale of random collision is given by
\begin{eqnarray}
\tau_{\rm coll}&=&\frac{500}{N^{2}}\left(\frac{R_{\rm H}}{\mbox{Mpc}}
\right)^{3}\left(\frac{r_{\rm gal}}{0.12\mbox{Mpc}}\right)^{-2}
\nonumber\\
&&\qquad\times\left(\frac{\sigma_{\rm gal}}{100\mbox{km~s}^{-1}}
\right)^{-4}\left(\frac{\sigma_{\rm halo}}{300\mbox{km~s}^{-1}}
\right)^{3}\mbox{Gyr},
\end{eqnarray}
where $N$ is the number of satellite galaxies, $r_{\rm gal}$ is their
radius, and $\sigma_{\rm halo}$ and $\sigma_{\rm gal}$ are the 1D
velocity dispersions of the common halo and satellite galaxies,
respectively \citep{MH97}.  With a probability of $\Delta t/\tau_{\rm
coll}$, where $\Delta t$ is the timestep corresponding to the redshift
interval $\Delta z$, a  satellite galaxy merges with another randomly
picked satellite. 

Consider the case that two galaxies of masses $m_1$ and $m_2 (>m_1)$
merge together.  If the mass ratio $f=m_1/m_2$ is larger than a certain
critical value of $f_{\rm bulge}$, we assume that a starburst occurs and
all the cold gas turns into stars and hot gas, which fills the halo, and
all of the stars populate the bulge of a new galaxy.  On the other hand, if
$f<f_{\rm bulge}$, no starburst occurs and a smaller galaxy is simply
absorbed into the disk of a larger galaxy.  These processes are repeated
until the output redshift.
We classify galaxies into different morphological types according to the
$B$-band bulge-to-disk luminosity ratio $B/D$.  In this paper, galaxies
with $B/D > 2/3$, and $B/D<2/3$ are classified as ellipticals/S0s and
spirals, respectively. This method of type classification well
reproduces the observed type mix.

\begin{figure}
  \begin{center}
    \FigureFile(120mm,80mm){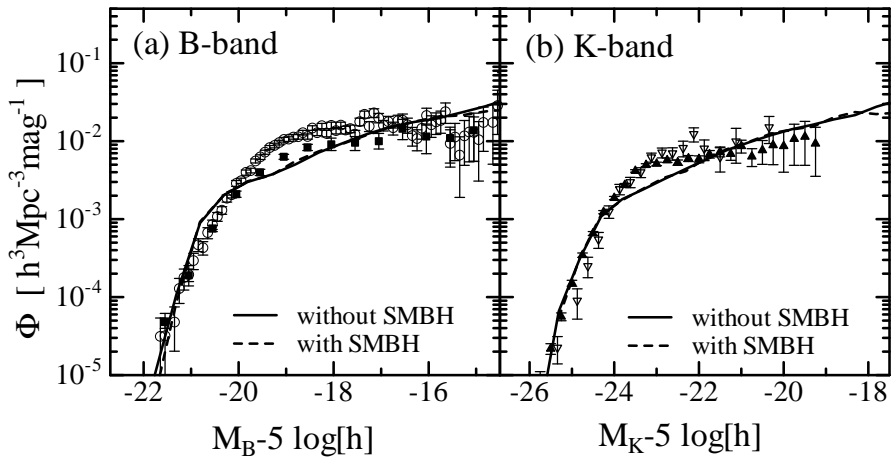}
  \end{center}
  \caption{Local luminosity functions in the (a) $B$-band and (b) $K$-band. 
  The thick line shows the result of the model without SMBH formation. The
  short dashed line shows the model with SMBH formation. Symbols with errorbars in (a)
 indicate the observational data from APM (\cite{APM}, {\it
 filled squares}) and 2dF (\cite{G2dF}, {\it open circles}). Symbols
 in (b) indicate the data from Gardner et al. (1997, {\it open inverted triangles}), and 2MASS (\cite{K3}, {\it filled triangles}).
  }\label{fig:gal-lum}
\end{figure}

The above procedure is a standard one in the SAM for galaxy formation.
Model parameters are determined by comparison with observations of the local
 Universe.
In this study,
we use the astrophysical parameters 
determined by \citet{NTGY01} from local observations such as 
luminosity functions, and galaxy number
counts in the Hubble Deep Field. The adopted parameters of
this model are tabulated in Table \ref{tab:astro}. In Figure \ref{fig:gal-lum}
 we plot the results of local luminosity functions of galaxies represented
by solid lines. Note that the resultant luminosity functions hardly
change if the SMBH formation model is included (dashed lines; see the
next section). Symbols with
errorbars indicate observational results from the $B$-band redshift
surveys (APM, \cite{APM}; 2dF, \cite{G2dF}) and from the $K$-band redshift surveys (\cite{K2}; 2MASS, \cite{K3}). 
 As can be seen, the results of our model
using these parameters are generally consistent with the observations, at
least with the APM result. 

\begin{table}
\caption{Model Parameters}  
\begin{center}
\label{tab:astro}
\begin{tabular}{ccccccccccccc}
\hline
\hline
 \multicolumn{4}{c}{cosmological parameters} &
& \multicolumn{6}{c}{astrophysical parameters} \\
\cline{1-4} \cline{6-11}
$\Omega_{0}$ & $\lambda_{0}$ &$h$ &$\sigma_{8}$ & &
$V_{\rm hot}$ (km~s$^{-1}$)
& $\alpha_{\rm hot}$ & $\tau_{*}^{0}$ (Gyr)
& $\alpha_*$ & $f_{\rm bulge}$ & $\Upsilon$\\
\hline
0.3&0.7&0.7& 1  && 280 & 2.5 & 1.5 & -2   & 0.5 & 1.5\\
\hline
\end{tabular}
\end{center}
\end{table}

\section{Model of Quasar Formation}\label{qsomodel}
\begin{figure}
  \begin{center}
    \FigureFile(70mm,70mm){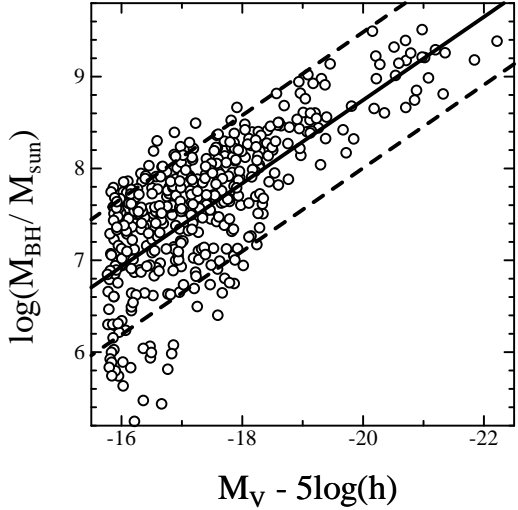}
  \end{center}
  \caption{The relation absolute $V$-band spheroid magnitude - mass of
 SMBH. The open circles are an absolute $V$-band magnitude limited sample of
 spheroids in our model. The thick solid line is the observational relation
 obtained by {\citet{Mag98}}. The dashed lines indicate the $1\sigma$
 scatter in the observations.} 
  \label{fig:bulge-bh}
\end{figure}

\begin{figure}
  \begin{center}
    \FigureFile(120mm,80mm){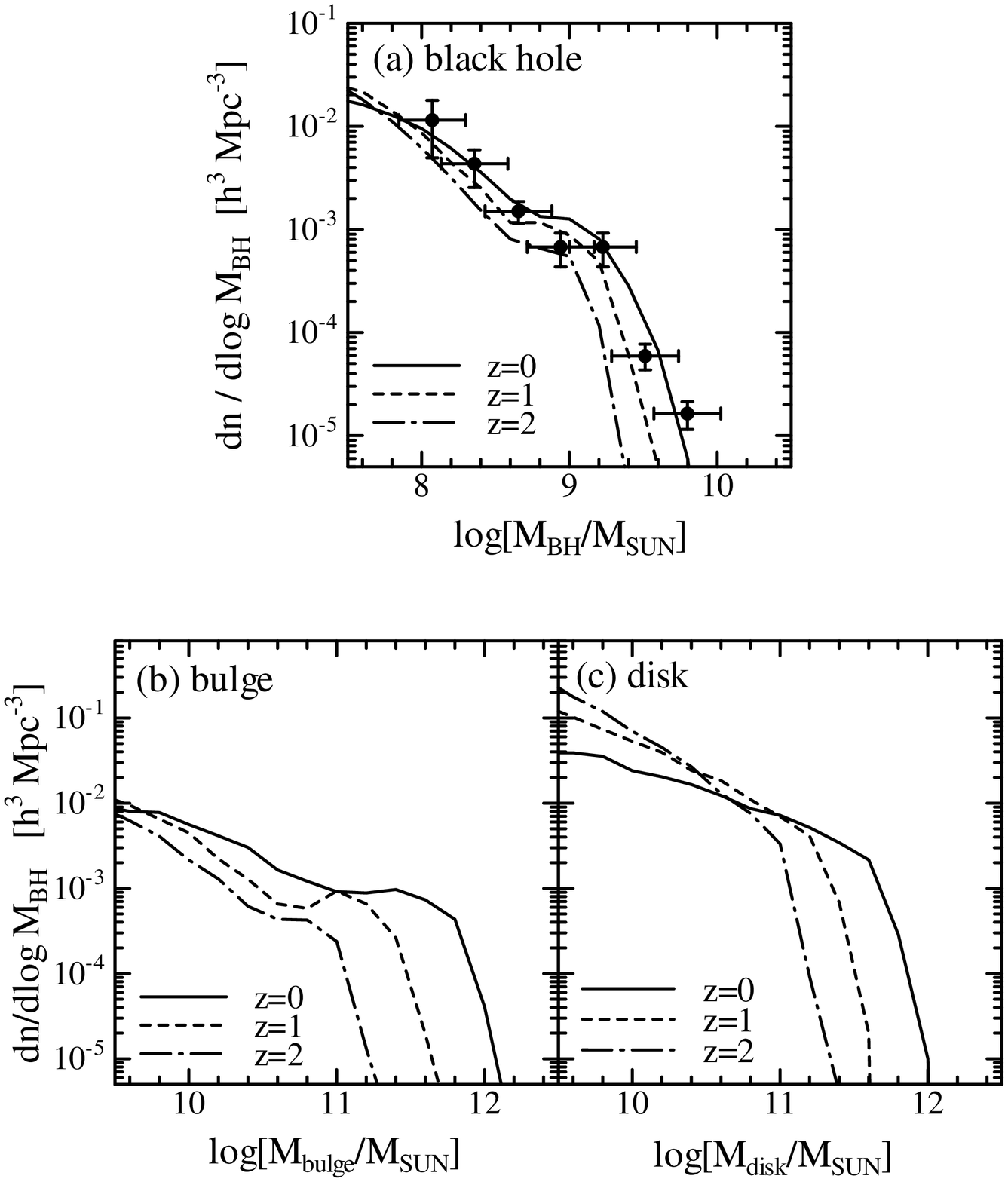}
  \end{center}
  \caption{(a) The black holes mass function of models for $f_{\rm BH} =0.03$ as a function of epoch. The solid, short-dashed and dot-dashed lines indicate
 the results at $z=0, 1$ and $2$ respectively. The symbols with
 errorbars are the
 present black hole mass function obtained by \citet{SSMD}. (b) The bulge
 and (c) the disk mass functions of model galaxies as a function of epoch. The solid, short-dashed and dot-dashed lines indicate
 the results at $z=0, 1$ and $2$ respectively.}
  \label{fig:bh-mass}
\end{figure}

In this section, we introduce a quasar formation and evolution model
into our SAM. 
As mentioned earlier, the masses of SMBHs have tight correlation with the spheroid masses of
 their host galaxies (e.g. \cite{KR95}; \cite{Mag98}; \cite{MF00})
 and the hosts of quasars  
found in the local Universe are giant elliptical galaxies or  galaxies
 displaying evidence of major mergers of galaxies (\cite{BKSS}). Moreover,
in SAMs for galaxy formation, it is assumed
that a galaxy-galaxy major merger leads to the formation of a spheroid.  
 Therefore, we assume that SMBHs grow by merging and are fueled by accreted
 cold  gas  during major mergers of galaxies.
 When host galaxies merge, pre-existing SMBHs
sink to the center of the new merged galaxy owing to dynamical friction and
finally coalesce. The timescale for this process is unknown, but for
the sake of simplicity  we assume that SMBHs merge instantaneously. 
Gas-dynamical simulations have demonstrated that the major merger of
 galaxies can
drive substantial gaseous inflows and trigger starburst activity
(\cite{NW83}; \cite{MH94}, \yearcite{MH96}; \cite{BH96}). Thus,
we assume that during major merger, some fraction of the cold gas
that is proportional to the total mass of stars newly formed at starburst is
accreted onto the newly formed SMBH. Under this assumption,
the mass of cold gas accreted on a SMBH is given by  
\begin{eqnarray}  
 M_{\rm acc} &=& f_{\rm BH} \Delta M_{*, \rm burst}, \label{eq:bhaccret}
\end{eqnarray} 
where $f_{\rm BH}$ is a constant and $\Delta M_{*, \rm burst} $ is the
total mass of stars formed at starburst. $\Delta M_{*, \rm burst} $ is
derived in the Appendix.  
The free parameter of $f_{\rm BH}$ is fixed 
by matching  the observed relation between a spheroid luminosity and
a black hole mass found by \citet{Mag98} and we find that the favorable value of
$f_{\rm BH}$ is nearly $0.03$. In Figure \ref{fig:bulge-bh} 
we show scatterplots (open circles) of the absolute $V$-band
magnitudes of spheroids versus  masses of SMBHs of model for $f_{\rm
BH}= 0.03$.  
The thick solid line is the observational relation  and the dashed lines
are the $1\sigma$  scatter in the observations obtained
by \citet{Mag98}. For $f_{\rm BH} \lesssim 0.3$, changing $f_{\rm BH}$
shifts the black hole mass almost linearly.
The obtained gas fraction  ($f_{\rm BH}=0.03$)  is so small that the inclusion of
SMBH formation  does 
not change the properties of galaxies in the local Universe. 
In Figure \ref{fig:gal-lum},
the dashed lines show the results of the model with the SMBH
formation. This result differs negligibly from the result of
the model without SMBH formation.
 Therefore, we use the same
astrophysical parameters tabulated in Table \ref{tab:astro}  regardless of
inclusion of the SMBH formation model.
Figure \ref{fig:bh-mass} (a) shows black hole mass functions in our
model at a series of redshifts. This indicates that the number density
of the most massive black holes increases monotonically with time in the
scenario where SMBHs grow by accretion of gas and by merging. In Figure
\ref{fig:bh-mass} (a), we superpose the present black hole mass function
obtained by \citet{SSMD}. They derived this black hole mass function
from the observed radio luminosity function of nearby radio-quiet
galaxies and  
the empirical correlation between radio luminosities from the nuclei of radio-quiet galaxies and the mass of their black holes. 
Our model result is consistent with their mass function.
For comparison, we also plot the mass functions of bulge and disk for
all galaxies in Figure \ref{fig:bh-mass} (b) and (c), respectively. The steep slopes at low masses
of mass functions of black hole and bulge are mainly due to random
collisions between satellite galaxies in this model.

To obtain the observed linear relation between a spheroid luminosity and
a black hole mass, Kauffmann and Haehnelt (\yearcite{KH00}) adopted  model of
fueling process in which the ratio of accreted mass to total available
 cold gas mass scales with halo circular velocity in the same way as
the mass of stars formed per unit mass of cooling gas. 
While their approach is similar to ours, their star formation and
feedback models are different from ours and they do not consider random
collisions.  Therefore, their resultant
model description is slightly differ from ours in equation (\ref{eq:bhaccret}).  

\begin{figure}
  \begin{center}
    \FigureFile(120mm,80mm){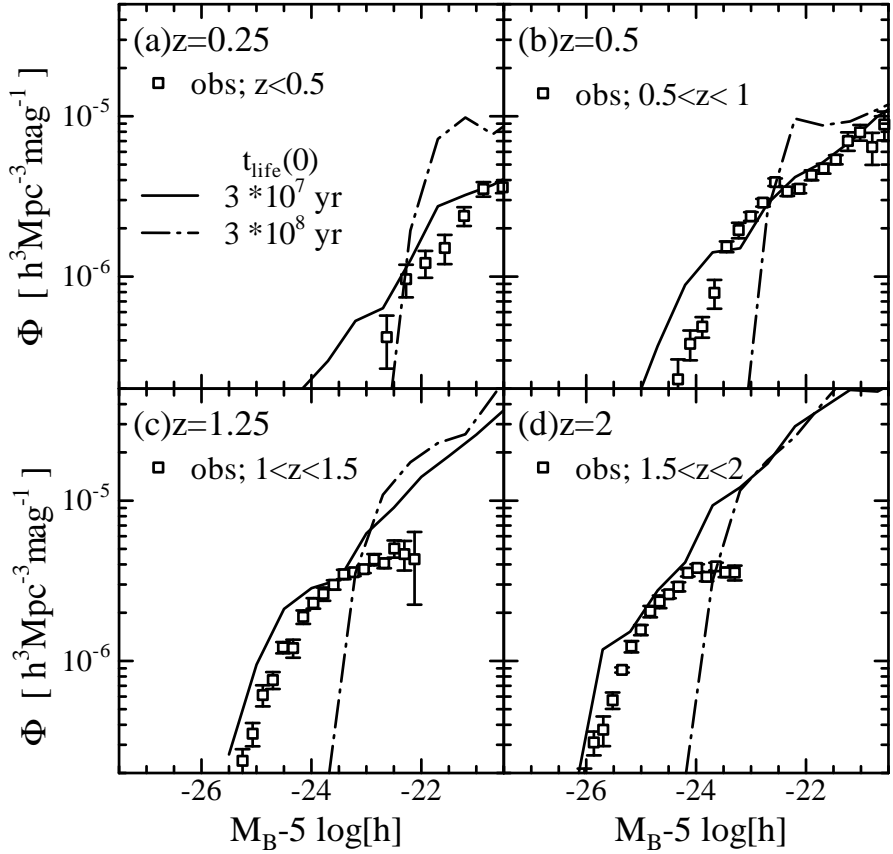}
  \end{center}
  \caption{The $B$-band quasar luminosity functions at (a)$z=0.25$, (b) $z=0.5$, (c)$z=1.25$
 and (d)$z=2.0$. The solid lines are $t_{\rm life}(0)=3 \times 10^7 {\rm yr}$ and the dot-dashed lines are $t_{\rm life}(0)=3 \times 10^8 {\rm yr}$. The symbols show results from the 2dF 10k catalogue (\cite{Cr01a}) reanalyzed by Takeuchi for a cosmology $\Omega_0=0.3, \lambda_0=0.7$ and $h=0.7$.} 
  \label{fig:qso-lum}
\end{figure}

Next, we consider the light curve of quasars. 
We assume that a fixed fraction of the rest mass energy of the accreted
gas is radiated in the $B$-band and the quasar life timescale  
$t_{\rm life}(z)$ scales with the dynamical time scale $t_{\rm dyn}$  of the
host galaxy where $t_{\rm dyn} \propto r_{\rm gal}/\sigma_{\rm gal} \propto
R_{\rm H}/V_{\rm circ}$.
Here we adopt the $B$-band
 luminosity of a quasar at time $t$ after the major merger as follows; 
\begin{equation} 
  L_B(t) =  L_B (\rm{peak}) \exp(-t/ t_{\rm life}) \label{eq:qso-lc}.
\end{equation} 
The peak luminosity $L_B (\rm{peak})$ is given by 
\begin{equation}
  L_B ({\rm peak}) =  \frac{\epsilon_B M_{\rm acc} c^2}{t_{\rm life}}, \label{eq:qso-peak} 
\end{equation}
where $\epsilon_B$ is the radiative efficiency in $B$-band, $t_{\rm
life}$ is the quasar life timescale and $c$ is the speed of light. 
 In order to determine the parameter
$\epsilon_B$ and the present quasar life timescale
 $t_{\rm life}(0)$,      
we have chosen them to match our model luminosity function with the
observed abundance of bright quasars at $z=2$. We obtain
$\epsilon_B = 0.005$ and $t_{\rm life}(0)= 3.0 \times 10^{7} {\rm
yr}$. The resulting luminosity functions at four different redshifts are
shown in Figure \ref{fig:qso-lum}. 
We superpose the luminosity functions derived from the 2dF 10k catalogue
(\cite{Cr01a})  for a cosmology with $\Omega_0=0.3,
\lambda_0=0.7$ and $h=0.7$, which is analyzed and kindly provided by T. T. Takeuchi. He used the method of \citet{EEP88} for the
estimation of the luminosity functions.
In order to reanalyze the error with greater accuracy, they applied bootstrap
resampling according to the method of \citet{TYI00}. Absolute $B$-band magnitudes were
derived for the quasars using the $k$-corrections derived by \citet{CV90}. 
Our model reproduces reasonably well the evolution of observed
luminosity functions. Thus, in the next section, we use these model
parameters  in order to
investigate the environments of quasars.

For comparison, we also plot the result of model with
$\epsilon_B = 0.005$ and $t_{\rm life}(0)= 3.0 \times 10^{8} {\rm yr}$
in Figure \ref{fig:qso-lum} (dot-dashed lines).
In this case, the abundance of luminous quasars decreases. 
To prolong a quasar life timescale affects the quasar luminosity
function due to the following two factors: a decrease in the peak
luminosity $L_{\rm B}$ (eq.[\ref{eq:qso-peak}]) and an increase in the exponential factor $\exp(-t/ t_{\rm life})$
in equation (\ref{eq:qso-lc}). 
 For the majority of bright quasars, 
 the elapsed time $t$ since the major merger is  much smaller than the quasar
 life timescale $t_{\rm  life}$, $t/t_{\rm life} \ll
 1$. Therefore, the former factor dominates the latter and the number of
 luminous quasars decreases.
 Thus, a long
quasar life timescale results in a very steep quasar luminosity function. Note
that if we change the radiative efficiency $\epsilon_B$, the quasar
luminosities simply scale by a constant factor in our model. Thus, changing
$\epsilon_B$ shifts the luminosity function horizontally.    

\section{Environments of Quasars}\label{env}
In this section, we investigate the environments of quasars using our model. 
We consider the halo mass dependence of the mean
 number of quasars per halo and the probability
 distribution of the number of galaxies around quasars
 as  characterizations of the 
 environments of quasars. This is because 
 the former is one of  measures of the
relation between quasars and dark matter distributions and the latter
reflects the relationship between galaxies and quasars.

\begin{figure}
  \begin{center}
    \FigureFile(120mm,80mm){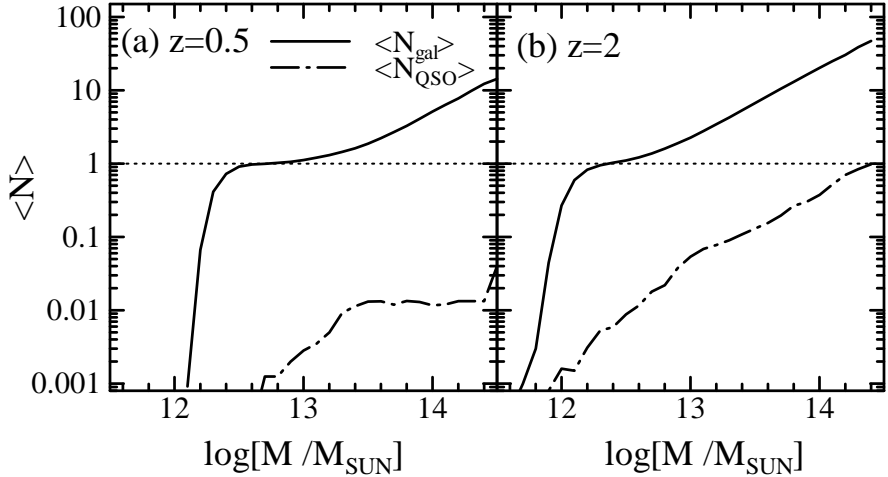}
  \end{center}
  \caption{The mean numbers of galaxies with $M_{\rm B}-5 \log(h) < -19$
 (solid lines)
 and quasars with $M_{\rm B}-5 \log(h) < -21$ (dot-dashed lines) per  halo with mass
  $M$ at (a) $z=0.5$ and (b) $z=2.0$. The horizontal dotted line marks
 $\langle N_{\rm gal}\rangle = 1$ and $\langle N_{\rm QSO}\rangle=1$} \label{fig:number-gal}
\end{figure}

In Figure \ref{fig:number-gal}, we plot $\langle N_{\rm gal}(M) \rangle$ and
$\langle N_{\rm QSO}(M)\rangle$ that denote the mean number of galaxies
and quasars per halo with mass $M$, respectively, at (a) $z=0.5$ and
(b) $z=2.0$. We select  galaxies with $M_{\rm B}-5 \log(h) < -19$ and
quasars with $M_{\rm B}-5 \log(h) < -21$, where $M_{{\rm B}}$ is absolute
$B$-band magnitude. 
It should be noted that changing the magnitude of selection criteria for
galaxies  and
quasars would alter these results, but qualitative features are not altered. 
As is seen in Figure \ref{fig:number-gal}, there are more galaxies and quasars at high $z$.
At higher redshift, halos have more cold gas available to form stars and
to fuel SMBHs because there has been relatively little time for star
formation to deplete the cold gas at these redshifts. Thus, the number of luminous galaxies grows. Furthermore, at
higher redshift,
both timescales of the dynamical friction and the random collisions are
shorter  because the mass density of a halo is higher. Therefore, the galaxy
merging rate increases.  Consequently, the number of quasars also
grows. Moreover, the decrease in the quasar life timescale $t_{\rm life}$
with redshift
 also contributes to  the increase in the number of quasars because quasars
 become brighter as a result of decrease in $t_{\rm life}$ 
(eq. [\ref{eq:qso-peak}]).

\begin{figure}
  \begin{center}
    \FigureFile(70mm,70mm){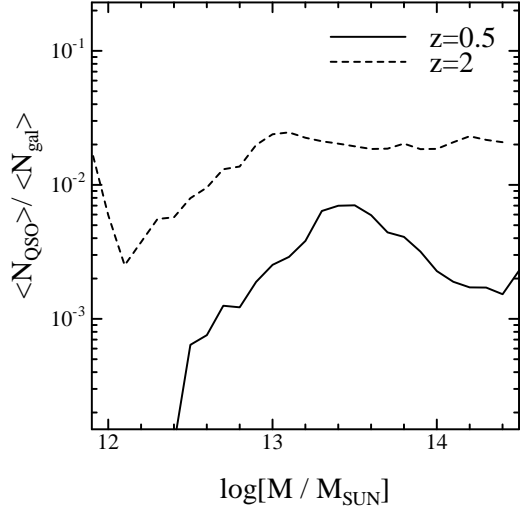}
  \end{center}
  \caption{The ratio of the mean number of galaxies with $M_{\rm B}-5 \log(h) < -19$ to the mean number of quasars with $M_{\rm B}-5 \log(h) < -21$ per halo with mass  $M$ at $z=0.5$ (solid line) and $z=2.0$ (dot-dashed line). } \label{fig:qg-ratio}
\end{figure}

\begin{figure}
  \begin{center}
    \FigureFile(70mm,70mm){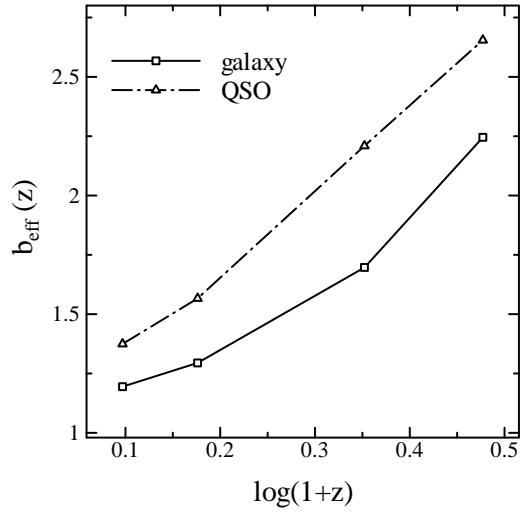}
  \end{center}
  \caption{The effective bias parameter of galaxies with $M_{\rm B}-5 \log(h) < -19$ (squares with solid line) and quasars with $M_{\rm B}-5 \log(h) < -21$ (triangles with dot-dashed line) at $z=0.25$, $z=0.5$, $z=1.25$ and $z=2.0$.} 
\label{fig:bias}
\end{figure}

From Figure \ref{fig:number-gal}, we find that the dependence of $\langle
N_{\rm QSO}(M) \rangle$ on  halo mass $M$ is different from the
dependence of $\langle N_{\rm gal}(M) \rangle$. Furthermore, Figure
\ref{fig:qg-ratio} shows that 
the ratio of  $\langle N_{\rm QSO}(M) \rangle$ to $\langle N_{\rm
gal}(M) \rangle$ varies with redshift and halo mass. 
 \citet{Ben00} used a combination of
cosmological $N$-body simulation and semi-analytic modeling
 of galaxy formation and
showed that the galaxy spatial distribution is sensitive to the
efficiency with which galaxies form in halos with different
mass. \citet{Sel00} also obtained the same conclusion using an analytic model of galaxy clustering. 
These results are applicable to the quasar spatial distribution. Therefore,   
our result indicates that the clustering properties of galaxies are
not the same as those of quasars and 
that the bias in the spatial distribution of galaxies relative to that of dark
matter is not the same as the bias in the spatial distribution of
quasars. Assumed that biases are independent of
 scale,  we can calculate effective biases using the method of
 \citet{Bau99} as follows;  
\begin{equation}
b_{\rm eff}(z) = 
 \frac{
 \int b(M,z) \langle N (M,z)\rangle n(M|z) {\rm d} M
 } 
 {
 \int \langle N (M,z) \rangle n(M|z) {\rm d} M
 },
 \label{eq:effbias}
\end{equation}
where $b(M,z)$ is the bias parameter
for dark matter halos of mass $M$ at $z$,
$\langle N(M,z) \rangle$ denotes the mean number of objects (galaxies or
quasars) in a halo of mass
$M$ at $z$ that satisfy the selection criteria and $n(M|z)$ is the dark
halo mass function at $z$. 
Our SAM adopts the Press-Schechter mass function 
which is given by
\begin{equation}
n(M|z)dM = \sqrt{\frac{2}{\pi}} \frac{\rho_0}{M} 
\frac{\delta_c(z)}{\sigma^{2}(M)}
\left|  \frac{d\sigma(M)}{dM} \right| 
\exp \left[-\frac{1}{2} \frac{\delta_c^{2}(z)}{\sigma^{2}(M)} \right] dM,
\label{eq:PSmass}
\end{equation}
where $\rho_0$ is the present mean density of the universe,
$\sigma(M)$ is the rms linear density fluctuation on the scale $M$ at
$z=0$ and
$\delta_c(z)=\delta_c/D(z)$. $D(z)$ is the linear growth factor,
normalized to unity at the present day and $\delta_{c}$ is the linear critical
density contrast at the collapse epoch.
Here, we use an approximate formula of
$\delta_c$ for spatially flat cosmological model (\cite{NS97}). 
The bias parameter for dark matter halos is given by \citet{Jing98};
\begin{equation}
b(M,z) = \left\{1 + \frac{1}{\delta_c} \left[
\frac{\delta^{2}_{c}(z)}{\sigma^{2}(M)} -1 \right] \right\}
 \left[ \frac{\sigma^{4}(M)}{2 \delta^{4}_{c}(z)} + 1 \right]^{(0.06 - 0.02n_{\rm eff})}, 
\label{eq:bias}
\end{equation}
where $n_{\rm eff}$ is the effective spectral index of the power spectrum, 
${\rm d}\ln P(k)/{\rm d}\ln k$,  at the wavenumber defined by the 
Lagrangian radius of the dark matter halo, $k = 2 \pi /r_{L}$ and $r_{L}
= \left(3M/4\pi \rho_{0}\right)^{1/3}$. Figure \ref{fig:bias} shows the
evolution of effective bias for galaxies with $M_{\rm B}-5 \log(h) < -19$
 and quasars with $M_{\rm B}-5 \log(h) < -21$. As is seen in Figure
 \ref{fig:bias}, quasars
 are higher biased tracer than galaxies. Furthermore, the evolution of quasar bias
 is different from that of galaxy bias. This reflects the difference in
 th dependence  on  halo mass $M$ and
 redshift of $\langle N_{\rm QSO}(M,z) \rangle$ and   
$\langle N_{\rm gal}(M,z) \rangle$. Note that these effective biases are
valid for large scale where objects (galaxies or quasars) which
contribute two-point correlation function populate different halos.    

\begin{figure}
  \begin{center}
    \FigureFile(120mm,80mm){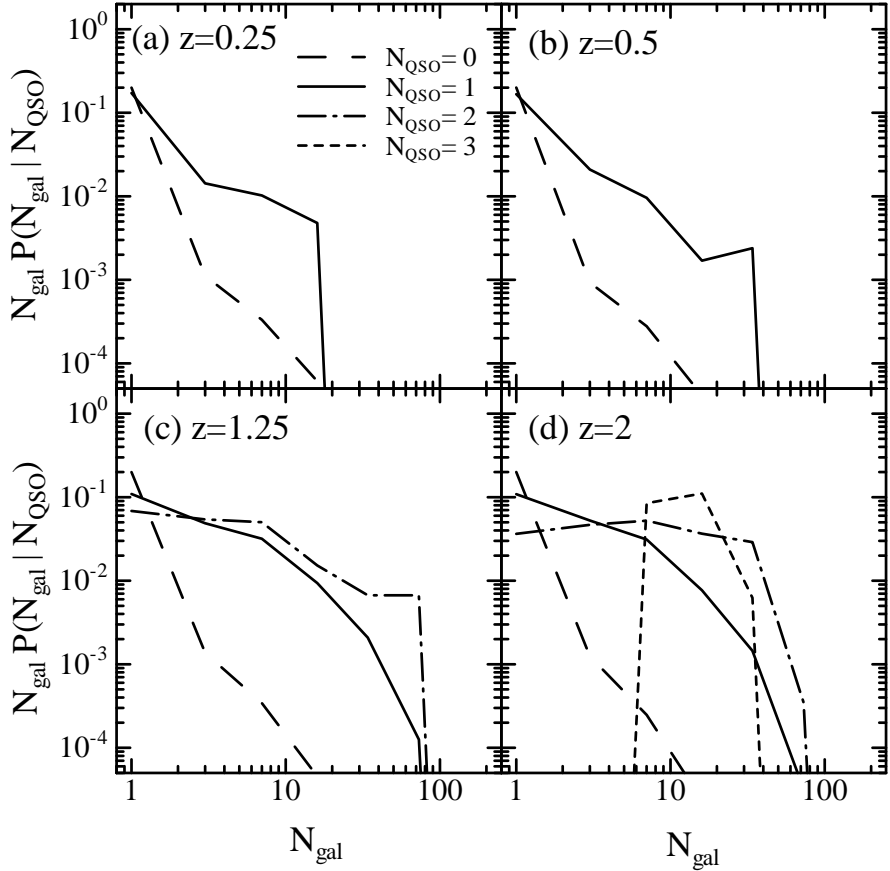}
  \end{center}
  \caption{The probability distribution for the numbers of
galaxies around quasars
 (a)$z=0.25$, (b) $z=0.5$, (c)$z=1.25$  and (d)$z=2.0$. The selected
 galaxies are brighter than $M_{\rm B}-5 \log(h) = -19$ and the selected
 quasars are brighter than $M_{\rm B}-5 \log(h) = -22$. Long-dashed, solid, dot-dashed
 and short-dashed lines show results for $N_{\rm QSO} = 0, 1, 2$ and $3$ respectively.} \label{fig:gnd}
\end{figure}

Next, we formulate the conditional probability that a halo with $N_{\rm
QSO}$ quasars has $N_{\rm gal}$ galaxies. 
The number density of the halos which contains $N_{\rm gal}$ galaxies
and $N_{\rm QSO}$ quasars at $z$ is obtained from the following
expression:   
\begin{equation}
n(N_{\rm gal},N_{\rm QSO}|z) = \int N(N_{\rm gal},N_{\rm QSO}|M,z) n(M|z) dM, \label{eq:ng-q}
\end{equation}
where $N(N_{\rm gal},N_{\rm QSO}|M,z) dN_{\rm gal} dN_{\rm QSO}$ denotes
the number of the halos with mass
$M$ which contains $N_{\rm gal} \sim N_{\rm gal}+dN_{\rm
gal}$ galaxies and $N_{\rm QSO}
\sim N_{\rm QSO}+dN_{\rm QSO}$ quasars at $z$ and $n(M|z)$ is the dark
halo mass function at $z$. 
The number density of the halos which contain $N_{\rm QSO}$ quasars at
$z$ is obtained from the following expression:
\begin{equation}
n(N_{\rm QSO}|z) = \int N(N_{\rm QSO}|M,z) n(M|z) dM, \label{eq:nq}
\end{equation}
where $N(N_{\rm QSO}|M,z) dN_{\rm QSO}$ denotes the number of the halos
with mass $M$ which contain $N_{\rm QSO} \sim N_{\rm QSO}+dN_{\rm QSO}$
quasars at $z$. 
From equation (\ref{eq:ng-q}) and (\ref{eq:nq}), the conditional
probability that the halo with $N_{\rm QSO}$ quasars has $N_{\rm gal}
\sim N_{\rm gal}+dN_{\rm gal}$ galaxies at $z$ is given by
\begin{equation}
P(N_{\rm gal}|N_{\rm QSO},z) dN_{\rm
gal} = \frac{n(N_{\rm gal},N_{\rm QSO}|z)}{n(N_{\rm QSO}|z)} dN_{\rm
gal}. \label{eq:pgq} 
\end{equation}
As is seen in the above formulation,
given $N(N_{\rm gal},N_{\rm QSO}|M,z)$ and $N(N_{\rm
QSO}|M,z)$ from the quasar formation model,  
one can calculate the probability distribution for the number of
galaxies around quasars.
Figure \ref{fig:gnd} shows these galaxy number
distribution functions around quasars estimated by  our
model. The results are shown for quasars
brighter than $M_{\rm B}-5 \log(h) = -22$ and for galaxies brighter than
$M_{\rm B}-5 \log(h) = -19$. Note that at $z=0.25$ and $z=0.5$ $P(N_{\rm gal}|N_{\rm QSO}=2)=0$ and $P(N_{\rm gal}|N_{\rm QSO}=3)=0$
for all $N_{\rm gal}$ (Fig. \ref{fig:gnd}(a) and (b)) and that at
$z=1.25$ $P(N_{\rm gal}|N_{\rm QSO}=3)=0$ for all $N_{\rm gal}$
(Fig. \ref{fig:gnd}(c)). At lower redshift, a halo has at most
one quasar. Fig \ref{fig:gnd}(a) and (b) show that the halo which has
one quasar contains several galaxies by high probability. These results indicate that most quasars tend to reside in groups of
galaxies at $0.2 \lesssim z \lesssim 0.5$ and is consistent with the
observation at $z \lesssim 0.4$ (e.g. \cite{BC91}; \cite{FBK96};
\cite{MD01}).
 On the other hand,
at higher redshift, the numbers of galaxies in the halo with one or
two quasars is from several to dozens (Fig \ref{fig:gnd}(c) and
(d)). These results indicate that
 quasars locate in ranging from small groups of galaxies to clusters of
galaxies. Thus at $1 \lesssim z \lesssim 2$ quasars seem to reside in
more varied environments than at lower redshift.  Kauffmann and Haehnelt
 (\yearcite{KH02})  used a combination of cosmological $N$-body
 simulation and semi-analytic modeling of galaxy and quasar formation,
 and showed that  the ratio of the amplitude of the quasar-galaxy  cross correlation function to that of the galaxy autocorrelation
 function decrease  with  redshift.
This indicates that the difference between galaxy and quasar
 distribution becomes smaller at higher redshift. Thus, our results 
 obtained by $P(N_{\rm gal}|N_{\rm QSO},z)$  is not
 in conflict with their results.      

\section{Summary and Discussion}\label{disc}
We have constructed a unified semi-analytic model for galaxy and quasar
formation and  have predicted the mean number of quasars
per halo with mass $M$, $\langle N_{\rm QSO}(M) \rangle$, the effective bias
 parameter of quasars $b_{{\rm eff, QSO}} (z)$ and
probability distribution of the number of galaxies around quasars,
$P(N_{\rm gal}|N_{\rm QSO})$, as characterizations of the environments of quasars.
These quantities reflect the processes of quasar formation such as the
amount of cold gas available for fueling, the galaxy merger rate and
the quasar life timescale. Therefore, by comparing these
predictions with observations, one  will be able to constrain quasar
formation models.  

Our model can reproduce not only general form of the galaxy luminosity
 functions in the local Universe but also the observed relation of the SMBH mass to spheroid  luminosity, and the quasar luminosity functions
at different  redshifts (Fig.\ref{fig:bulge-bh} and 
Fig.\ref{fig:qso-lum}). 
Using this model, we have
 shown $\langle N_{\rm QSO}(M) \rangle$ and $P(N_{\rm gal}|N_{\rm QSO})$.
The ratio of $\langle N_{\rm QSO}(M) \rangle$ to $\langle N_{\rm gal}(M) \rangle$ varies with halo mass
in our model (Fig\ref{fig:number-gal}). These results
 of our model suggest that the clustering of galaxies is not the same as
 the clustering of quasars and the effective bias parameter of quasars and its
 evolution are different from these of galaxies (Fig.\ref{fig:bias}). 
Furthermore, we predict  the galaxy number
distribution function around quasars, $P(N_{\rm gal}|N_{\rm QSO})$  
(Fig\ref{fig:gnd}). At lower
redshifts ($0.2 \lesssim z \lesssim 0.5$), most halos which have quasars
have at most several galaxies. 
This indicates that most quasars reside in 
groups of galaxies.  
On the other hand, at higher redshift ($1 \lesssim z \lesssim
2$), the number of galaxies in the halo with  quasars is from several to 
dozens; quasars reside in ranging from small groups of galaxies to
 clusters of galaxies. These results show that most quasars at
higher redshift reside in more varied environments than at lower redshift. This
model prediction is checkable by statistics of galaxies
around quasars which will be obtained in future.

It is  still controversial whether the environments of quasars depend on
their optical and radio luminosities. Some authors 
 have
claimed that radio-loud quasars were located in richer environments than
radio-quiet quasars at at $z<0.6$ (e.g. \cite{YG84}; \cite{YG87}; \cite{EYG91};
\cite{HRV91}).
 However,
other people obtained a different result. For example, \citet{HCJ95}
observed the galaxy environment of radio-loud quasars and radio-quiet quasars
and concluded that there is no significant difference in the
richness. Recent studies support this conclusion (e.g. \cite{WLLS01}). 
The discrepancies between different studies may be caused partly by 
too small quasar samples and by differences in sample selection of
quasars.
This situation will soon improve with the availability of a new generation
of very large quasar surveys such as the 2dF quasar redshift survey (\cite{Cr01a}) and the
Sloan Digital Sky Survey (\cite{SDSS}). Although we do not deal with radio
properties of quasars in this paper, our investigation of quasar
environments will also provide a clue for understanding the radio character of quasar environments. 

The mean number of quasars
per halo, $\langle N_{\rm QSO}(M) \rangle$, and
probability distribution of the number of galaxies around quasars,
$P(N_{\rm gal}|N_{\rm QSO})$, used in this study can provide
some useful features of the quasar environments. 
Furthermore, the spatial galaxy-quasar correlation function is
used in order to quantify the galaxy environments around a quasar.
Therefore, for the further investigation of environments and clustering
of quasars and in order to constrain the quasar formation model, it is also
necessary to predict  spatial distribution of galaxies and quasars.
We will show the results using  the combination of
cosmological $N$-body simulation and SAM for formation of galaxy and
quasar in the near future.

%%%\acknowledgments
\bigskip
We would like to thank T. T. Takeuchi for providing us with the reanalyzed data of the
quasar luminosity functions derived from the 2dF 10k catalogue.
We are also grateful to K. Okoshi, H. Yahagi and S. Yoshioka for useful
comments and discussions. We also thank to the anonymous referee for a
thorough reading of the manuscript and for his
valuable suggestions and comments, which improved our paper very much. 
Numerical computations in this work were partly carried out at the
Astronomical Data Analysis Center of the National Astronomical
Observatory, Japan. This work has been supported in part by the
Grant-in-Aid for the Scientific Research Funds (13640249) of the
Ministry of Education, Culture, Sports, Science and Technology of Japan.

\appendix
\section*{Star Formation and Gas Evolution}\label{sec:ap1}
In this appendix, we summarize our model of star formation and gas
evolution. We use a simple
instantaneous recycling approximation of model star formation, feedback
and chemical enrichment.
The following  
difference equations describe the evolution of the mass of cold
gas $M_{\rm cold}$, hot gas $M_{\rm hot}$, and long lived stars $M_{\rm
star}$ at each time step.
\begin{eqnarray}
\dot{M}_{\rm cold} &=&  -\dot{M}_{*} + R \dot{M}_{*} -\beta \dot{M}_{*}, 
 \label{eq:coldeq} \\
\dot{M}_{\rm hot} &=&\beta \dot{M}_{*}, \label{eq:hoteq} \\
\dot{M}_{\rm star} &=& \dot{M}_{*} - R \dot{M}_{*},  \label{eq:stareq} 
\end{eqnarray}
where $\dot{M}_{*}={M_{\rm cold}}/{\tau_{*}}$ is star formation rate,
$R$ is the gas fraction returned by evolved stars, and $\beta$ is the
efficiency of reheating. In this paper, $R=0.25$.
The solutions of these equations are the following:
\begin{eqnarray}
M_{\rm cold} &=& M_{\rm cold}^{0} \exp \left[-\left(1-R+\beta
  \right)\frac{t}{\tau_{*}} \right], \label{eq:coldsol} \\ 
M_{\rm hot} &=& M_{\rm hot}^{0} + \beta \Delta M_{*}, \label{eq:hotsol} \\
M_{\rm star} &=& M_{\rm star}^{0} + (1-R)\Delta M_{*}, \label{eq:starsol} 
\end{eqnarray}
where $M_{\rm cold}^{0}, M_{\rm hot}^{0}$ and $M_{\rm star}^{0}$ are the
masses of cold gas, hot gas and  long-lived stars from the previous time step, $t$ is the time sine the start
of the time
step, and $\Delta M_{*} = (M_{\rm
cold}^{0} - M_{\rm cold})/(1-R+\beta)$ is the mass of total
formed stars.    

When a starburst occurs, stars are formed in a very short
timescale. Thus, the starburst corresponds to $\tau_{*}/t \to 0$ in the above
solutions. In this case, the changes of masses are given by
\begin{eqnarray} 
M_{\rm cold} &=& 0, \label{eq:coldburst} \\ 
M_{\rm hot} &=& M_{\rm hot}^{0} +  \frac{\beta M_{\rm cold}^{0}}{1-R+\beta}, \label{eq:hotburst}  \\
M_{\rm star} &=& M_{\rm star}^{0} + 
 \frac{(1-R) M_{\rm cold}^{0}}{1-R+\beta} \label{eq:starburst}  
\end{eqnarray}
 and the total star mass formed at starburst becomes 
\begin{equation}
\Delta M_{*, \rm burst} = \frac{M_{\rm cold}^{0}}{1-R+\beta}. \label{eq:totstar} 
\end{equation}
From equation (\ref{eq:totstar}), we can obtain the mass of accreted cold
gas onto a black hole (eq.[\ref{eq:bhaccret}]).

\end{document}